\documentclass[11pt,a4paper]{article}
\pdfoutput=1
\usepackage{jcappub}

\usepackage{mathrsfs}
\usepackage{bm}
\usepackage{subfigure}
\usepackage{feynmp}
\usepackage{extarrows}
\usepackage{slashed}
\usepackage{dcolumn}
\usepackage{verbatim}
\usepackage{here}
\usepackage{multirow} 
\usepackage{xcolor}
\allowdisplaybreaks[4]

\numberwithin{equation}{section}

\newcommand{\FR}[2]{\displaystyle\frac{\,{#1}\,}{#2}}
\newcommand{\fr}[2]{\mbox{$\frac{\,{#1}\,}{#2}$}}
\newcommand{\n}{\nonumber}

\def\bge{\begin{equation}}
\def\ede{\end{equation}}
\def\bga{\begin{aligned}}
\def\eda{\end{aligned}}
\def\bgp{\begin{pmatrix}}
\def\edp{\end{pmatrix}}
\def\bgs{\begin{subequations}}
\def\eds{\end{subequations}}
\newcommand{\beq}{\begin{equation}}
\newcommand{\eeq}{\end{equation}}
\newcommand{\bq}{\begin{equation}}
\newcommand{\eq}{\end{equation}}
\newcommand{\ba}{\begin{array}}
\newcommand{\ea}{\end{array}}
\newcommand{\beqa}{\begin{eqnarray}}
\newcommand{\eeqa}{\end{eqnarray}}
\newcommand{\beqs}{\begin{subequations}}
\newcommand{\eeqs}{\end{subequations}}

\newcommand{\order}[1]{\mathcal{O}({#1})}
\def\di{{\mathrm{d}}}

\def\[{\left[}
\def\]{\right]}
\def\({\left(}
\def\){\right)}

\def\ra{\rightarrow}

\def\dis{\displaystyle}

\def\over{\overline}

\def\al{\alpha}
\def\be{\beta}

\def\la{\lambda}

\def\ep{\epsilon}
\def\lam{\lambda}

\def\si{\sigma}

\def\ZZ{\mathbb{Z}_2^{}}
\def\RR{\mathcal{R}}
\def\TT{\mathcal{T}}
\def\SS{\mathcal{S}}
\def\di{\mathrm{d}}

\def\pd{\partial}
\def\ld{{\mathscr{L}}}

\def\la{\langle}\def\ra{\rangle}

\def\to{\rightarrow}

\def\ii{\mathrm{i}}

\def\mut{\mu_*^{}}

\def\mT{m_{\TT}^{}}
\def\mTT{m_{\TT}^2}
\def\mt{m_t^{}}
\def\mtt{m_t^2}
\def\mS{m_{\mathcal{S}}^{}}

\def\Mp{M_{\text{Pl}^{}}}

\def\cutIF{\Lambda_{\text{INF}}^{}}

\def\End{\end{document}}

\title{\huge Extending Higgs Inflation with \\ TeV Scale New Physics}

\author[a,c,d]{\large Hong-Jian He,\,}
\author[a,b]{\large~~Zhong-Zhi Xianyu\,}

\affiliation[a\,]{Institute of Modern Physics and Center for High Energy Physics,\\
                  Tsinghua University, Beijing 100084, China}
\affiliation[b\,]{Theoretical Particle Physics and Cosmology Group, Department of Physics,\\
                \,King's College London, London WC2R 2LS, UK}
\affiliation[c\,]{Center for High Energy Physics, Peking University, Beijing 100871, China}
\affiliation[d\,]{Kavli Institute for Theoretical Physics China, CAS, Beijing 100190, China}

\emailAdd{hjhe@tsinghua.edu.cn, xianyuzhongzhi@gmail.com}

\abstract{
\\[1mm]
Higgs inflation is among the most economical and predictive inflation models,
although the original Higgs inflation requires tuning the Higgs or top mass
away from its current experimental value by more than $\,2\sigma$\, deviations,
and generally gives a negligible tensor-to-scalar ratio $\,r \sim 10^{-3}\,$ (if away from
the vicinity of critical point).  In this work, we construct a minimal extension of Higgs inflation,
by adding only two new weak-singlet particles at TeV scale,
a vector-quark $\,\TT\,$ and a real scalar $\,\SS\,$.\, The presence of singlets $\,(\TT,\,\SS )$\,
significantly impact the renormalization group running of the Higgs boson self-coupling.
With this, our model provides a wider range of the tensor-to-scalar ratio
$\,r=\order{0.1-10^{-3}}\,$,\,
consistent with the favored $\,r\,$ values by either BICEP2 or Planck data, while keeping
the successful prediction of the spectral index $\,n_s^{}\simeq 0.96\,$.\,
It further allows the Higgs and top masses to fully fit the collider measurements.
We also discuss implications for searching the predicted TeV-scale
vector-quark $\,\TT\,$ and scalar $\,\SS\,$ at the LHC and future high energy $pp$ colliders.
}

\keywords{\\[1mm]
Inflation, Cosmology of theories beyond the SM, Particle physics\,$-$\,Cosmology connection
\\[5mm]
JCAP (2014), final version [\,arXiv:1405.7331\,]
}

\begin{document}

\maketitle

\setlength{\baselineskip}{18pt}

\setcounter{page}{2}
\vspace*{10mm}
\section{Introduction}
\label{sec:1}
\vspace*{2mm}

 Cosmic inflation \cite{inf0,inf1,inf2,inf3,inf3a} successfully describes the evolution of the very early universe.
 It not only resolves problems of the standard big-bang cosmology (such as horizon problem
 and flatness problem), but also generates primordial fluctuations, which shape the large scale structure
 of the universe. The amplitudes and shapes of these fluctuations can be directly tested by the observations
 of CMB and large scale structure.
 The recent BICEP2 discovery of $B$-mode in the CMB polarization \cite{BICEP2}
 has provided strong evidence that the gravitational waves (with a quantum origin)
 were created during inflation.
 The BICEP2 tensor-to-scalar ratio $\,r = 0.20^{+0.07}_{-0.05}\,$
 \cite{BICEP2}, if confirmed,
 will strongly constrain viable models of inflation.
 With Planck normalization $\,(V/\ep )^{1/4}=0.027\Mp^{}\,$ \cite{PlanckCosPara},
 it also reveals the energy scale of inflation,
 $\,\cutIF = V^{1/4}_{}\simeq 2 \!\times\! 10^{16}\,$GeV,\,
 where $\,\ep = r/16\,$ is the first slow-roll parameter, and
 $\,\Mp^{}\simeq 2.44\!\times\! 10^{18}\,$GeV is the reduced Planck mass.
 The current BICEP2 data have some tension with the indirect measurement of $\,r\,$ by Planck satellite
 \cite{Planck} if one assumes a negligible running $\,\al_s^{} (=\text{d}n_s^{}/\text{d}\ln k\,$) of spectrum index
 $\,n_s^{}\,$.\,  Although a large $\,\al_s^{}\,$ can reconcile both Planck and BICEP2 results, such
 a large running would probably ruin the slow-roll approximation. It is desirable to further verify
 the dust model adopted by BICEP2 and quantify other possible sources of foreground
 contamination such as the magnetized dust associated with radio loops\,\cite{dust1}
 or the potential dust polarization effect\,\cite{dust2} allowed
 by the current Planck data\,\cite{Planck}\cite{Planck-new}.
 The upcoming data from Planck, Keck Array and other $B$-mode measurements
 (such as ABS, ACTPol, EBEX, POLARBEAR, Spider and SPT) are expected to further pin down the situation.

 Among all the existing inflation models, the Higgs inflation \cite{HI1,HI2,HIF-rev}
 appears the most economical and predictive one since it is the only model
 which uses a discovered particle --- the Higgs boson\,\cite{LHC2013}, as the inflaton.
 The original model of Higgs inflation makes use of the unique dimension-4 operator of
 the Higgs-gravity interaction, $\,\xi\mathcal{R}H^\dag H$\,,\,
 where $\,\RR\,$ is the Ricci scalar curvature and $\,H\,$ is the Higgs doublet.
 By taking $\,\xi\simeq 17000$\, as required by the observed strength
 of curvature perturbation, the model predicts at the tree-level that the spectral index
 $\,n_s^{}\simeq 0.967\,$ and the tensor-to-scalar ratio $\,r\simeq 0.0031\,$,\,
 which are in agreement with Planck data \cite{Planck}.
 When taking account of renormalization group (RG) running,
 the model also exhibits some tension\,\cite{Allison} with the measured masses
 of the Higgs boson and top quark \cite{LHC2013,mt-exp,mt-new}.

 It is clear that the original Higgs inflation model with large
 $\,\xi = \order{10^4}\,$ is disfavoured by the BICEP2 data.
 But there are still parameter regions in this model to give a sizable $\,r\,$.\,
 As noticed in Refs.\,\cite{Allison,HIcrit1,HIcrit2}, when the Higgs mass reaches its ``critical point'',
 with which the Higgs self-coupling $\,\lambda\,$ runs to vanishingly small value at inflation scale,
 a successful inflation could be achieved by only a mildly large $\,\xi = \order{10}\,$,\,
 while the tensor-to-scalar ratio $\,r\,$ can be as large as $\order{0.1}$.
 A relatively small $\,\xi\,$ also makes the Higgs inflation fully free from the potential problem of
 unitarity violation \cite{HIF-rev,XRH,Burgess}.
 But, it is known\,\cite{Allison,HIcrit1,HIcrit2,Spencer}
 that the critical point for Higgs mass or top-quark mass
 as required in this case is already outside the $\,2\si\,$ ranges of the collider measurements,
 $\,m_h^{} = 125.6 \pm 0.2(\text{stat}) \pm 0.3(\text{syst})\,$GeV \cite{LHC2013}
 and $\,m_t^{} = 173.39^{+1.12}_{-0.98}\,$GeV \cite{mt-new}\,\footnote{The current most precise measurement
 on the top-quark mass is given by the world combination of
 the ATLAS, CMS, CDF and D0 experiments\,\cite{mt-exp},
 $\,m_t^{} = 173.34 \pm 0.27(\text{stat}) \pm 0.71(\text{syst})\,$GeV. This is based on
 the best fit to the mass parameter as implemented in the respective Monte Carlo (MC) code for generating
 the theory input, and is usually called MC top mass.
 It is shown\,\cite{mt-new} that this MC mass definition can be converted to a theoretically well-defined
 short-distance mass definition with an uncertainty of
 $\,\sim\! 1\,\text{GeV}$,\, and the resultant pole mass is,
 $\,m_t^{} = 173.39^{+1.12}_{-0.98}\,$GeV \cite{mt-new}.
 As a further note, the Snowmass study\,\cite{snowmass2013-top} found
 that the upgraded high luminosity LHC can eventually reduce the error $\,\Delta m_t^{}\,$ down to $500$\,MeV,
 and an $e^+e^-$ lepton collider (such as the ILC) can measure $\,\Delta m_t^{}\,$ to $100$\,MeV level.}.\,
 If we use the current central values of Higgs mass and top quark mass,
 together with the two-loop RG running,
 the Higgs self-coupling will turn negative around $\,10^{11}$\,GeV,
 which is far below the expected inflation scale.

  But, we should keep in mind that all the analyses mentioned above are based on a strong assumption
  that the standard model (SM) holds all the way up to the inflation scale, and no new particle
  beyond the SM would exist. This is unlikely the case, since there are many motivations for
  new physics showing up at the TeV scale.
  Hence, it is intriguing to study how the presence of new particles will improve the original
  Higgs inflation model and provide the corresponding new discovery signatures at colliders.
  Some interesting attempts appeared lately \cite{ExtHI}.

  In this paper, we construct a new minimal extension for the Higgs inflation, by adding
  two weak-singlet particles at the TeV scale, a vector-quark $\,\TT\,$ and a real scalar $\,\SS\,$.\,
  We will demonstrate that within such a minimal setting, the model can perfectly fit both
  the cosmology data (including BICEP2\,\cite{BICEP2})
  and the collider measurements on Higgs and top masses \cite{LHC2013,mt-exp,mt-new}.
  This provides an effective way to remove the tension between the original Higgs inflation model and
  collider data, as well as allowing a wide range of the tensor-to-scalar ratio $\,r\,$
  in light of the BICEP2\,\cite{BICEP2} and Planck data\,\cite{Planck}.
  We also note that some particle phenomenology of singlet vector-quarks were
  studied before in very different contexts \cite{vectorF1,vectorF2}.

\vspace*{4mm}
\section{Minimal Extension of Higgs Inflation with Two Weak-Singlets}
\label{sec:2}
\vspace*{2mm}

  We may naively expect that the minimal extension of the original Higgs inflation\,\cite{HI1}
  would be to add only one new particle, presumably a scalar. However, as we will show below,
  the current experimental values of the SM Higgs and top masses lie in such a critical region that
  any new heavy particle interacting with the SM Higgs would strongly affect the qualitative RG-running
  behavior of the Higgs self-coupling. Thus, adding just one new particle would require
  a higher degree of fine-tuning in general \cite{EllisRoss}.

  To make this clear, let us inspect the one-loop $\beta$-function of Higgs self-coupling
  $\,\lam\,$ in the SM,
  \bge
   \be_\lam^{(\text{SM})}
   \,=\, \FR{1}{(4\pi)^2}
   \Big\{24\lam^2-6y_t^2+\FR{3}{8}\big[2g^4+(g^2+g'^2)^2\big]+\lam(-9g^2-3g'^2+12y_t^2)\Big\},
  \ede
 where $\,y_t^{}\,$ is the top Yukawa coupling, and $(g,\,g')$
 are gauge couplings of $SU(2)_L^{}\otimes U(1)_Y^{}$.\,
 Here we have neglected the small Yukawa couplings of all light SM fermions, expect that of top quark.
 The contributions from gauge couplings are also numerically small as compared to that of
 $\,\lam\,$ and $\,y_t^{}\,$.\, Hence, the running of $\,\lam\,$ are largely determined
 by the two competing factors, $\,\lam\,$ and $\,y_t^{}\,$,
 which can be further expressed in terms of Higgs mass $\,m_h^{}\,$ and top mass $\,m_t^{}\,$
 through $\,\lam = m_h^2/2v^2\,$ and $\,y_t^{}=\sqrt{2}m_t^{}/v\,$ at tree-level,
 where $\, v\simeq 246$\,GeV is the vacuum expectation value of the Higgs field.
 Due to the large top Yukawa coupling $\,y_t^{}\simeq 1\,$,\, this $\beta$-function will decrease
 $\,\lam\,$ with the increasing energy, and finally pushes $\,\lam$ to zero and negative values
 at a scale $\,\mut\,$ which we call the turning point.
 Inputting the current experimental central values $\,m_h^{}=125.6$\,GeV and $\,m_t^{}=173.3$\,GeV,
 and using the two-loop RG running, we find that the turning point is around $\,\mut \sim 10^{11}$\,GeV.
 This is far below the inflation scale $\,\Lambda_{\text{INF}}^{}\simeq 2.3\times 10^{16}$\,GeV,
 as inferred from BICEP2 measurement\,\cite{BICEP2}.
 But the location of this turning point is rather sensitive to the Higgs and top masses.
 For instance, if we input a smaller top mass $\,m_t^{}\simeq 171\,$GeV
 (beyond the $2\sigma$ lower bound of $m_t^{}$ data\,\cite{mt-new}),
 the turning point $\,\mut\,$ will be quickly shifted to the Planck scale
 $\,\Mp\simeq 2.44\times 10^{18}\,$GeV.
 In fact, this is the main observation invoked in the recent ``critical point scenario''
 \cite{Allison,HIcrit1,HIcrit2}  of Higgs inflation as mentioned in Sec.\,\ref{sec:1}.

 When adding new particles coupled to the Higgs field, the qualitative picture is the same as before:
 bosons and fermions will drive $\,\lam\,$ towards positive and negative values, respectively.
 Thus, adding just one scalar or fermion will generally destroy the art of exquisite balancing.
 Hence, we are naturally led to construct a minimal viable extension of the SM Higgs inflation
 by adding just two new weak-singlets, a real scalar $\,\SS\,$ and a vector-quark $\,\TT\,$.
 In the following, we will demonstrate that this model can provide a successful Higgs inflation,
 and achieve full agreements with the current experimental data from both cosmology and colliders.

 In our construction, we impose a simple $\ZZ$ symmetry under which $\,(\SS,\,\TT_R^{})\,$ and
 $\,(t,\,b)$\, are $\ZZ$-odd, while all other fields are $\ZZ$-even.
 In Table\,\ref{tab:1}, we summarize the quantum number assignments for the
 third family quarks, the Higgs doublet, and the new fields $\,(\TT,\,\SS)\,$
 under $\,SU(2)_L^{}\otimes U(1)_Y^{}\otimes \ZZ\,$.\, Here we have defined,
 $\,Q_{3L}^{}=(t,\,b)_L^T\,$.\,
 All other fields have the same assignments as in the SM.

 \begin{table}
 \begin{center}
 \begin{tabular}{c||ccc|cc|cc}
   \hline\hline
   &&&&&&&
   \\[-3mm]
   Group &  $Q_{3L}^{}$  &  $t_R^{}$  &  $b_R^{}$  &  $\TT_L^{}$ & $\TT_R^{}$  & $H$ & $\SS$
   \\[-3mm]
   &&&&&&&
   \\ \hline
   &&&&&&&
   \\[-3mm]
   $SU(2)_L^{}$  & \bf{2} &  1 & 1 & 1 & 1 & \bf{2} & 1
   \\[-3mm]
   &&&&&&&
   \\
   $U(1)_Y^{}$  & $\fr{1}{6}$ & $\fr{2}{3}$ & $-\fr{1}{3}$ & $\fr{2}{3}$ & $\fr{2}{3}$ & $\fr{1}{2}$ & 0
   \\[-3mm]
   &&&&&&&
   \\
   $\ZZ$ & $-$ & $-$ & $-$ & $+$ & $-$ & $+$ & $-$
   \\[-3mm]
   &&&&&&&
   \\
   \hline\hline
 \end{tabular}
  \caption{Quantum number assignments for the third family quarks $(t,\,b)$, the Higgs doublet,
          and the new fields $\,(\SS,\,\TT)\,$
          under $\,SU(2)_L^{}\otimes U(1)_Y^{}\otimes \ZZ\,$.
          All other fields have the same assignments as in the SM and are $\ZZ$ even.
          Here we have defined, $\,Q_{3L}^{}=(t,\,b)_L^T\,$.\,}
 \label{tab:1}
 \end{center}
 \end{table}

 Thus, we can write down the general scalar potential for the Higgs doublet
 $\,H$\, and the real singlet $\,\SS\,$ as follows,
 \beqa
   \label{ScalarPotential}
   V(H,\SS) ~=\,
   -\mu_1^2 H^\dag H - \fr{1}{2}\mu_2^2{\SS}^2 + \lam_1^{}(H^\dag H)^2
   +\fr{1}{4}\lam_2^{}{\SS}^4+\fr{1}{2}\lam_3^{}{\SS}^2H^\dag H
   + \kappa\, \SS \,,
 \eeqa
 where $\,H=(\pi^+,\,\fr{1}{\sqrt 2}(v+h+\ii\pi^0))$,\, with $\,v\simeq 246$\,GeV
 being the electroweak vacuum expectation value of the Higgs field.
 The quadratic term of $\,\SS\,$ has a negative mass-term, and
 thus the $\,\ZZ\,$ symmetry is spontaneously broken by the nonzero vacuum expectation value (VEV)
 of $\,\SS\,$ field, $\,\la {\SS}\ra = u =\order{\text{TeV}} \gg v\,$.
 In our construction, we have conjectured that all interactions are $\ZZ$ symmetric.
 Thus, any possible soft $\ZZ$ breaking term has to be noninteracting, and
 the last term of \eqref{ScalarPotential} gives the unique soft $\ZZ$ breaking term.
 This term lifts the degenerate vacua of $\,\SS$\,
 and avoids the domain wall problem\,\cite{DW} associated with the spontaneous $\ZZ$ breaking.
 Requiring the potential \eqref{ScalarPotential} to be asymptotically bounded from below,
 we have the tree-level conditions,
 $\,\lam_1^{},\lam_2^{}>0\,$ and
 $\,\lam_1^{}\lam_2^{}>\fr{1}{4}\lam_3^2\,$.\,
 Minimizing the scalar potential \eqref{ScalarPotential}, we derive two extremal conditions,
 \beqs
 \beqa
 &&
 \lam_1^{}v^2 +\frac{1}{2}\lam_3^{}u^2 ~=~ \mu_1^2 \,,
 \\[1mm]
 &&
 \frac{1}{2}\lam_3^{}v^2 +\lam_2^{}u^2 + \frac{\,\kappa\,}{\,u\,} ~=~ \mu_2^2 \,,
 \eeqa
 \eeqs
 In practice, the soft breaking term is small,  $\,\kappa/(v^3,u^3) \ll 1$\,.\,
 For instance, besides $\,v\simeq 246\,$GeV as fixed by the Fermi constant,
 we have $\,u=\order{\text{TeV}}\,$
 and $\,\kappa =\order{1\!-\!10\,\text{GeV}}^3\,$.\,
 So, we can treat the $\,\kappa\,$ term as a perturbation
 and only keep linear terms in $\,\kappa\,$.
 Thus, we can derive the VEVs,
 \beqs
 \beqa
 &&
 v ~\equiv~ v_0^{}+\delta_v^{}\,, ~~~~~
 u ~\equiv~ u_0^{}+\delta_u^{}\,,
 \\[2mm]
 &&
 v_0^2 ~= \frac{\,2(2\lam_2^{}\mu_1^2 - \lam_3^{}\mu_2^2)\,}
                {\,4\lam_1^{}\lam_2^{}-\lam_3^2\,} \,, ~~~~~
 u_0^2 ~= \frac{\,2(2\lam_1^{}\mu_2^2 - \lam_3^{}\mu_1^2)\,}
                {\,4\lam_1^{}\lam_2^{}-\lam_3^2\,} \,,
 \\[0.5mm]
 &&
 \delta_v^{} ~\simeq
 \frac{\lam_3^{}}{\,4\lam_1^{}\lam_2^{}-\lam_3^2\,}
 \frac{\kappa}{v_0^{}u_0^{}} \,, ~~~~~
 \delta_u^{} ~\simeq
 \frac{\lam_1^{}\kappa}{\,\lam_3^{}\mu_1^2 -2\lam_1^{}\mu_2^2\,} \,.
 \eeqa
 \eeqs
 For our later numerical analysis of Higgs inflation in Sec.\,\ref{sec:3},
 we find that the small linear $\kappa$ term has negligible effect
 on our samples in Table\,\ref{tab:2}, because
 it does not affect the interaction terms and RG running.

 From Table\,\ref{tab:1}, we further construct the relevant Yukawa interactions for
 the $\,(t,\,\TT)\,$ sector,
 \beqa
 \label{eq:L-Yukawa}
   \ld^{}_{t\TT} \,=\,      -y_1^{} \bar{Q}_{3L}^{} \tilde H t_{R}^{}
              -y_2^{} \bar{Q}_{3L}^{}\tilde H \TT_R^{}
              - \frac{y_3^{}}{\sqrt{2}\,}{\SS}\,\bar\TT_L^{}\TT_R^{}
              - \frac{y_4^{}}{\sqrt{2}\,}\SS\,\bar{\TT}_L^{}t_R^{}
              + \text{h.c.} \,,
 \eeqa
 where $\,\tilde H =i\tau_2^{}H^*\,$ is the charge-conjugate of Higgs doublet.
 There is also a Yukawa term for $b$-quark mass generation, $\bar{Q}_{3L}^{} H b_{R}^{}$\,.\,
 We did not display this in \eqref{eq:L-Yukawa} since $b$-quark has no mixing with $\,\TT$\,.\,
 We note that the last term on the right-hand-side of \eqref{eq:L-Yukawa}
 is not independent and it can always be absorbed by a field redefinition,
 $\,\TT_R^{} \to \TT_R^{} - (y_4^{}/y_3^{})t_R^{}\,$.\,
 Hence, we will drop the last term of Eq.\,\eqref{eq:L-Yukawa},
 and only deal with the three independent
 couplings $(y_1^{},\,y_2^{},\,y_3^{})$ hereafter.
 It is clear that the $\ZZ$ symmetry singles out the $\,(t,\,\TT)\,$ sector,
 and disallows mixings between $\,(t,\,\TT)\,$ and the light up-type quarks in the first two families.
 Such Yukawa mixing terms could arise via effective dimension-5 operators, e.g.,
 $\,\bar{Q}_{jL}^{} \tilde{H} t_{R}^{}\SS/\Lambda_S^{}\,$ and
 $\,\bar{Q}_{jL}^{} \tilde{H} \TT_{R}^{}\SS/\Lambda_S^{}\,$,\,
 where the family index $\,j=1,2\,$ and $\,\Lambda_S^{}\,$ is the associated cutoff.
 This naturally explains why the mixings of the third family quarks with the first two families
 are much smaller than those among the first two families themselves, as indicated in the CKM matrix.

 After the spontaneous symmetry breaking, the scalars \,$(h,\,\SS)$\,
 form a $2\times 2$ mass-matrix $\,\mathbb{M}^2_{s}\,$ and its diagonalization gives
 the mass-eigenvalues $(m_h^{2},\,m_{\SS}^{2})$,
 \beqs
 \beqa
 &&
 \mathbb{M}^2_{s} \,=
 \begin{pmatrix}
 \lam_1v^2 & \lam_3 vu
 \\[1mm]
 \lam_3 vu & \lam_2 u^2
 \end{pmatrix}
 =\,
 U(\al)
 \begin{pmatrix}
 m_h^2 & 0
 \\[1mm]
 0     & m_{\SS}^2
 \end{pmatrix}\!
 U(\al)^T,
 \hspace*{6mm}
 \\[2mm]
 &&
  U(\al) \,=
 \begin{pmatrix}
 ~\,\cos\al & \sin\al
 \\[1mm]
 -\sin\al & \cos\al
 \end{pmatrix} \!,
  \hspace*{6mm}
 \tan\al ~=\, \frac{\,\lam_1^{} x^2 \!- \lam_2^{}z_h^2\,}{\,\lam_3^{} x (1 + z_h^2)\,} \,,
 \hspace*{15mm}
 \\[2mm]
 &&
 (m_h^2,\,m_{\SS}^2) \,=\, \dis
 \frac{u^2}{2}\!\left\{
 (\lam_1^{}x^2\!+\lam_2^{})\mp \left[(\lam_1^{}x^2\!-\lam_2)^2\!+4\lam_3^2x^2\,\right]^{\frac{1}{2}}
 \right\} \!,
 \eeqa
 \eeqs
 where we have defined the VEV ratio  $\,x\equiv v/u \ll 1\,$
 and the mass-ratio $\,z_h^{}\equiv m_h^{}/m_{\SS}^{}\,$.\,
 The orthogonal diagonalization matrix $\,U(\al)\,$ connects the weak-eigenbasis
 $(h,\,\SS)$ to the mass-eigenbasis $(\hat{h},\,\hat{\SS})$.\,
 For convenience, we will simply use the notations $(h,\,\SS)$ for mass-eigenstates
 in the following, unless specified otherwise.
 From these, we can further resolve the quartic scalar couplings in terms of
 the mass-eigenvalues and mixing angle,
 \bge
 \begin{aligned}
   &\lam_1^{} ~=~ \FR{\,m_h^2\cos^2\al + m_S^2\sin^2\al\,}{2v^2} \,, \\
   &\lam_2^{} ~=~ \FR{\,m_h^2\sin^2\al + m_S^2\cos^2\al\,}{2u^2} \,, \\
   &\lam_3^{} ~=~ \FR{\,m_S^2-m_h^2\,}{2vu} \sin 2\al \,.
 \end{aligned}
 \ede

 In parallel, for the fermion sector,
 we derive the following mass-matrix for the $\,(t,\,\TT)\,$,
 \beqa
 \mathbb{M}_{f}^{} \,= \,\frac{v}{\sqrt{2}\,}
 \begin{pmatrix}
 \,y_1^{} & y_2^{}
 \\[2mm]
 0 & y_3^{}x^{-1}
 \end{pmatrix}\!,
 \eeqa
 Then, we can diagonalize the symmetric matrix $\,\mathbb{M}_{f}^{}\mathbb{M}_{f}^{\dag}\,$
 by the left-handed rotation $\,U(\theta)\,$ from $(t,\,\TT)_L^T$ into the mass-eigenstates
 $(\hat{t},\,\hat{\TT})_L^T$,\, i.e.,
 $\,(t,\,\TT)_L^T=U(\theta)(\hat{t},\,\hat{\TT})_L^T\,$.\, Hence, we have
 \beqs
 \beqa
 \mathbb{M}_{f}^{}\mathbb{M}_{f}^{\dag} &\!\!=\!\!\!&
 U(\theta)
 \begin{pmatrix}
 \mtt & 0
 \\[1.5mm]
 0    & \mTT
 \end{pmatrix}
 U(\theta)^{\dag} \,,
 \\[2mm]
 U(\theta) &\!\!=\!\!\!&
 \begin{pmatrix}
 ~\,\cos\theta & \sin\theta
 \\[1mm]
 -\sin\theta & \cos\theta
 \end{pmatrix} ,
 \hspace*{8mm}
 \tan\theta ~=~ \frac{~x^2(y_1^2\!+\!y_2^2)-z_t^2y_3^2~}{x(1\!+\!z_t^2)y_2^{}y_3^{}} \,,
 \\[2mm]
 (m_t^2,\,m_{\TT}^2) &\!\!=\!\!\!&
 \frac{\,u^2}{4}\left[x^2(y_1^2\!+y_2^2) + y_3^2\right]\!
 \left\{ 1\mp \left[1-\frac{4x^2y_2^2y_3^2}{[x^2(y_1^2\!+y_2^2)+y_3^2]^2}\right]^{\!\frac{1}{2}}
 \right\} ,
 \hspace*{8mm}
 \eeqa
 \eeqs
 where we have defined the mass ratio $\,z_t^{} \equiv m_t^{}/m_{\TT}^{} \ll 1\,$.\,
 For convenience, we will simply denote the mass-eigenstates by the notations $(t,\,\TT)$
 in the following, unless specified otherwise.
 With these, we can resolve the Yukawa couplings $\,(y_1^{},\,y_2^{},\,y_3^{})$\,
 as functions of the quark mass-eigenvalues and the left-handed mixing angle,
 \bge
 \begin{aligned}
   y_1^{} &\,=\, \FR{\sqrt{2}\,z_t^{}}{\,v\(z_t^2\sin^2\theta+\cos^2\theta\)^{\frac{1}{2}}\,} \,,
   \\[1mm]
   y_2^{} &\,=\, \FR{( 1 - z_t^2 )\sin 2\theta}
                    {\,\sqrt{2}\,v\(z_t^2\sin^2\theta + \cos^2\theta\)^{\frac{1}{2}}\,} \,,
   \\[1mm]
   y_3^{} &\,=\, \FR{\,\sqrt{2}\,\mT\,}{u}\(z_t^2\sin^2\theta + \cos^2\theta\)^{\frac{1}{2}}_{} .
 \end{aligned}
 \ede

 For our construction, we include the unique dimension-4 operator as in the original Higgs inflation,
 \bge
 \label{eq:NMC}
   \Delta\ld_{\text{NMC}}^{} ~=~ \sqrt{-g\,}\,\xi\,\mathcal{R}H^\dag H \,,
 \ede
 where $\,\xi\,$ is the dimensionless non-minimal coupling\footnote{An alternative construction of
 Higgs inflation with Higgs boson minimally coupled to gravity is recently given
 in Ref.\,\cite{Xianyu:2014eba},
 where Einstein general relativity exhibits asymptotic safety in the ultraviolet region.}
 between the Ricci scalar-curvature $\,\mathcal{R}$\,  and the Higgs doublet \,$H$\,.\,
 In principle, there is another nonminimal term, $\,\xi_s^{}\mathcal{R}{S}^2$\,,\,
 for the new singlet scalar $\,{S}\,$.\,
 We do not add it here because it is irrelevant to our present study.
 The new interactions generated by the non-minimal term $\,\mathcal{R}H^\dag H$\,
 were extensively studied in Ref.\,\cite{XRH}.

\vspace*{2mm}
\section{Improved Scalar Potential and New Predictions for Higgs Inflation}
\label{sec:3}

 In this section, we systematically study the scalar potential by including the radiative corrections.
 With this, we can derive predictions on the inflationary observables, and compare them with the
 cosmology measurements including Planck\,\cite{Planck} and BICEP2\,\cite{BICEP2}.
 We will identify the parameter space of our minimal extension, which can fit well with
 the favored tensor-to-scalar ratio by the BICEP2 or Planck, as well as the collider data
 on Higgs and top masses.

\begin{figure}
   \centering
  \includegraphics[height=6.5cm,width=0.49\textwidth]{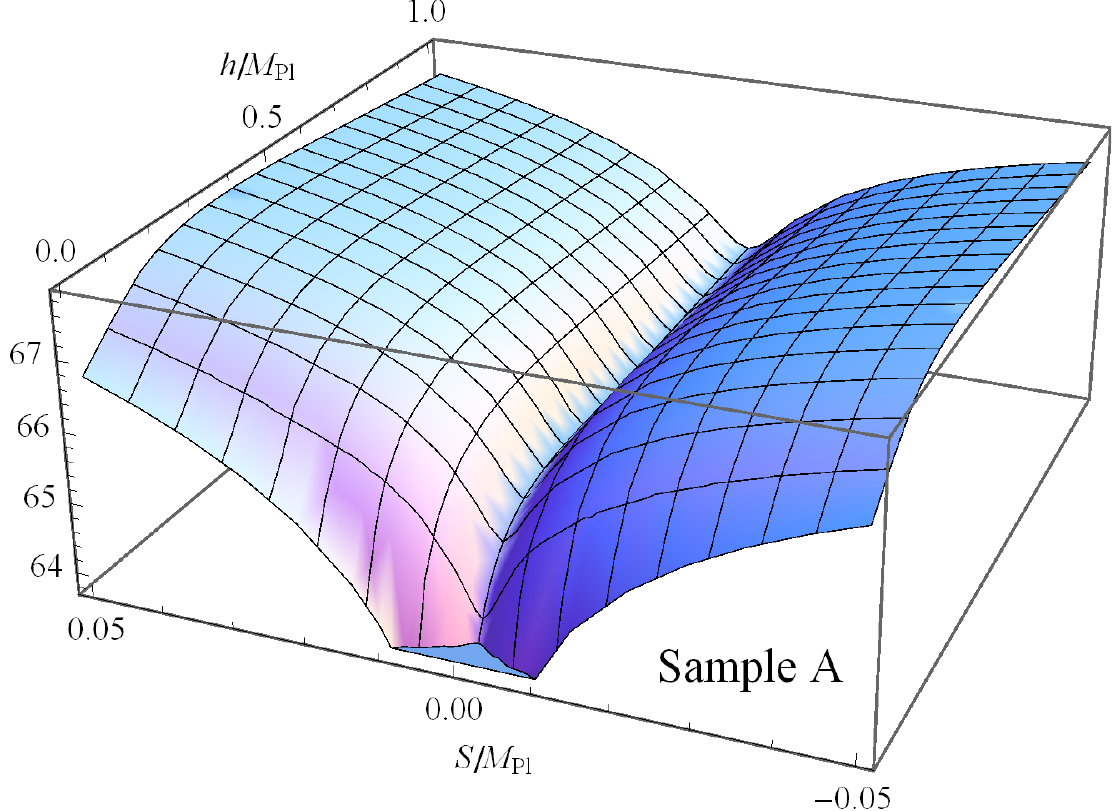}
  \includegraphics[height=6.5cm,width=0.49\textwidth]{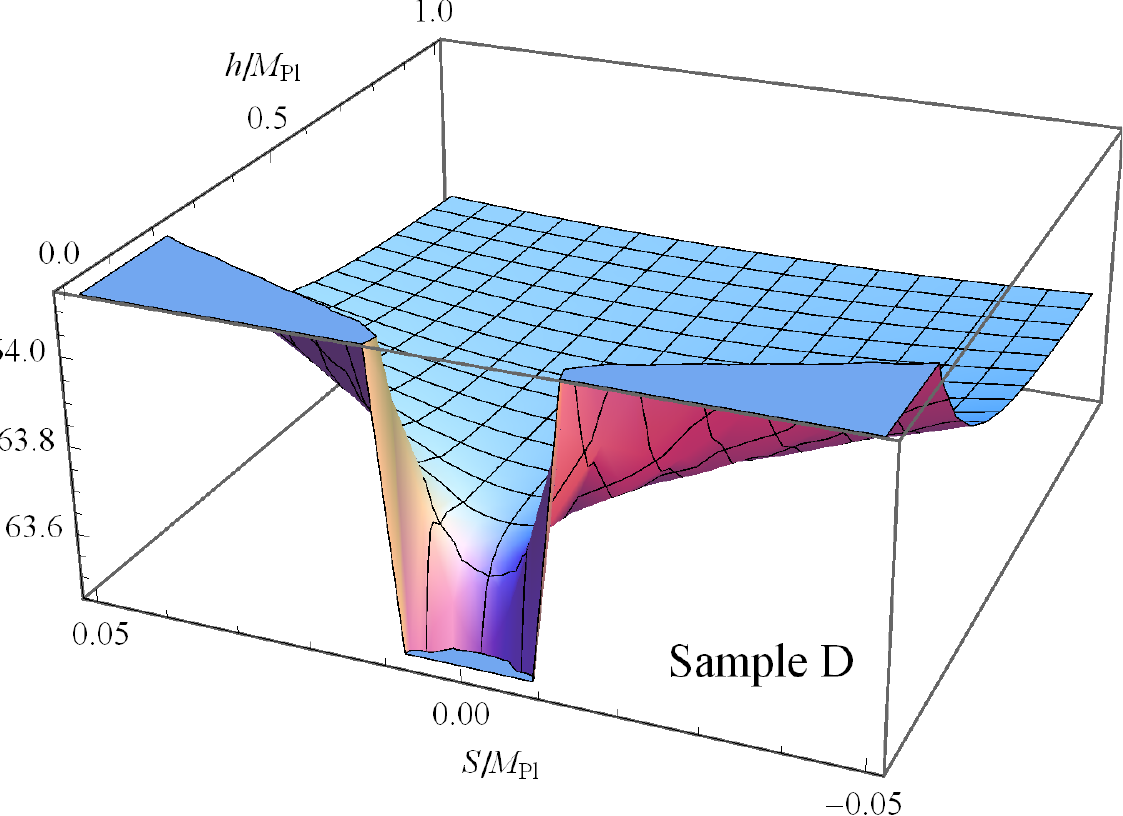}
   \caption{Scalar potential $\,V(h,\SS)$\, as a function of fields $\,h\,$ and $\,\SS\,$
            for Sample-A (left plot) and Sample-D (right plot).
            For the vertical axis in each plot, we depict the potential  $\,V(h,\SS)$\,
            in terms of $\,\log_{10}^{}(V/\text{GeV}^4)\,$.\,
            All inputs are taken from Table\,\ref{tab:2}.}
   \label{fig:3}
 \end{figure}

 The scalar sector of the model consists of a Higgs doublet $\,H\,$ and a real singlet $\,\SS\,$.\,
 At the inflation scale, the scalar potential is a function of
 the module $\,|H|$\,,\, where the four components of $\,H\,$ appear in the same manner.\,
 Without losing generality, we can choose
 $\,|H|=\fr{1}{\,\sqrt{2}\,}h$\, for simplicity \cite{Hinfd}.
 We first set $\,\SS =0\,$ for the inflation analysis,
 and the effect of nonzero $\,\SS\,$ will be readily included later (cf.\ Fig.\,\ref{fig:3}).
 With this setup, the scalar potential (\ref{ScalarPotential}) depends only on the Higgs field
 $\,h\,$,\, which is identified as the inflaton.

 Due to the presence of nonminimal coupling term $\xi RH^\dag H$ in \eqref{eq:NMC},
 the equation of motion for the spacetime metric $\,g_{\mu\nu}^{(J)}\,$
 differs from the Einstein equation of general relativity.
 This is conventionally called the Jordan frame, as marked by the superscript $(J)$
 of the metric. To analyze the inflation based on the standard slow-roll formulation,
 we will make the field-redefinition, $\,g_{\mu\nu}^{(J)}=\Omega^2 g_{\mu\nu}^{(E)}$,\,
 where $\,\Omega^2=1+\xi h^2/\Mp^2$\,.\, The metric $\,g_{\mu\nu}^{(E)}$\,
 defines the Einstein frame and takes the standard form of Friedmann-Robertson-Walker.
 After this transformation to Einstein frame,
 the kinetic term and the potential for $\,h\,$ becomes,
 \bge
 \ld_h^{} ~=~
 \sqrt{-g^{(E)}}\left[
 \FR{\Omega^2\!+6\xi^2h^2/\Mp^2}{2\Omega^4}(\pd_\mu h)^2-\FR{\lam h^4}{4\Omega^4}
 \right],
 \ede
 where we have ignored the VEV $v$ of Higgs field since it is negligible during inflation.
 We further make a field redefinition $\,\chi=\chi(h)\,$ such that,
 \beqa
   \FR{\di\chi}{\di h} \,=\, \FR{\,\(\Omega^2+6\xi^2h^2/\Mp^2\)^{\!\frac{1}{2}}_{}\,}{\Omega^2} \,.
 \eeqa
 Thus, the $\,\chi\,$ field is canonically normalized in Einstein frame.

 After including radiative corrections, the Higgs potential $\,V\,$
 can be compactly summarized as follows,
 \beqa
  \label{EffPotential}
   V \,=\,\FR{\lam(\mu)\, h^4}{~\left[1+\xi(\mu)h^2/\Mp^2\right]^2_{}~} \,,
 \eeqa
 where $\,\lam(\mu)\,$ and $\,\xi(\mu)\,$ are running couplings,
 which are inferred by solving the renormalization group equations.
 As argued in \cite{Bezrukov:2009db}, the beta functions for running couplings
 should be gauge-invariant and do not contain the anomalous dimension of
 Higgs field $\,h\,$.\footnote{We also note that the effect of the Higgs anomalous dimension
 is generally negligible and does not cause any visible effect in our numerical analysis.}
 The full set of beta functions for our analysis is presented in Appendix\,\ref{App}.
 Here we only highlight the difference of the beta functions in our model from that of the SM.

 The most important differences come from the new scalar $\,\SS\,$ and new vector-quark $\,\TT\,$.\,
 This not only introduces new couplings $\,\lam_{2,3}^{}\,$ and $\,y_{2,3}^{}\,$,\,
 but also modifies the $\beta$-functions of all relevant SM couplings,
 including the Higgs self-coupling $\,\lam_1^{}\,$ and the top-Yukawa coupling $\,y_1^{}\,$,\,
 as well as the three gauge couplings $(g_3^{},\,g,\,g')$.
 Besides, the nonminimal coupling $\,\xi\,$ also modifies $\beta$-functions
 through its correction to the Higgs field in the loop.
 This means that all the loop-lines of Higgs field $\,h\,$ should be multiplied
 by the factor $\,s\,$,
 \beqa
   s(h) ~=~ \FR{\Omega^2(h)}{\,\Omega^2(h)+6\xi^2h^2/\Mp^2\,} \,.
 \eeqa
 The net effect of this modification is to insert the proper $s$-factors
 in the corresponding terms in $\beta$-functions.
 The SM $\beta$-functions and anomalous dimension with appropriate $s$-insertion
 were given in Refs.\,\cite{Allison,sInsertion}.

 At this stage, there is a potential ambiguity in choosing the renormalization scale
 $\,\mu\,$ \cite{Bezrukov:2009db,Allison}.
 In the Einstein frame approach (denoted as prescription-I in the literature),
 the optimal choice is $\,\mu=h/\Omega(h)\,$,
 while in the Jordan frame approach (known as prescription-II in literature)
 the renormalization scale is chosen to be $\,\mu=h$\,.\,
 These two choices may be essentially different and could be regarded as the low energy remnants
 of different UV completions. Some recent studies attempted to reconcile these apparent differences,
 which suggests the quantum equivalence of the two frames \cite{FrameEqv}.
 We keep open-minded on this issue.
 For the current study, we will use the prescription-I \cite{Bezrukov:2009db},
 i.e., we work in Einstein frame and set the renormalization scale $\,\mu=h/\Omega(h)$\,.\,
 We also note that in a more sophisticated study, one could write $\,\mu= \kappa h/\Omega(h)\,$
 and adjust $\,\kappa\sim\order{1}$\, to minimize the loop-corrections to the effective potential \cite{Allison}.
 For simplicity, we will follow Ref.\,\cite{Allison} and set $\,\kappa =1\,$ in the current numerical analysis.
 When the radiative correction is dominated by top-loop, it may be natural to choose the renormalization scale
 $\,\mu= \kappa h/\Omega(h)\,$ with $\,\kappa=y_t/\!\sqrt{2}$\, instead.
 We note that ignoring such a factor $\,\kappa\,$ could cause an uncertainty in the choice of $\,\mu\,$
 and thus the numerics, though it is expected to be generally small.
 For illustration, let us take Sample-A in Table\,\ref{tab:2} as an example, and check how it may change
 by setting $\,\kappa=y_t/\!\sqrt{2}$\,.\, After a systematical analysis, we obtain the following
 new Sample-A$'$,
 \beq
 \ba{l}
  \hspace*{-3mm}
   (u,\,\mS,\,\mT) = (7,\, 2.87,\, 2.87)\,\text{TeV},~~~
   (\alpha,\,\theta) = (1.8,\, 1.319375)\!\times\! 10^{-2},~~~
   \xi = 7.815 \,.~~~
 \ea
 \eeq
 This is to be compared with the original Sample-A in Table\,\ref{tab:2} under $\kappa=1$\,.\,
 From this comparison, we see that the values of $\,u\,$ and $\,\alpha\,$ in Sample-A$'$
 remain the same as in Sample-A, while the values of $\,(\theta,\,\xi)\,$ change by about $(1\!-\!4)\%$, and
 the masses of $({\cal S},\,{\cal T} )$ by about $7\%$\,.\,
 Such small changes have little effect on the phenomenology,
 and our main conclusions remain the same.

 \begin{table}
 \centering
 \begin{tabular}{c|c|c|c|c|c|c}
   \hline\hline
   &&&&&&
   \\[-3mm]
   Sample &  $u$  &  $\mS$   & $\mT$   & $\alpha$          & $\theta$                & $\xi$
   \\
   &&&&&&
   \\[-3mm]
   \hline
   &&&&&&
   \\[-3mm]
    A     &  7\,TeV &  3.08\,TeV & 3.08\,TeV & $1.8\!\times\!10^{-2}$ & $1.33682\!\times\!10^{-2}$ & 7.53035
    \\[1.5mm]
    B     &  4\,TeV &  1.34\,TeV & 1.34\,TeV & $4.0\!\times\!10^{-2}$ & $3.00017\!\times\!10^{-2}$ & 10.464
    \\[1.5mm]
    C     &  4\,TeV &  1.288\,TeV & 1.288\,TeV & $4.0\!\times\!10^{-2}$ & $2.9898\!\times\!10^{-2}$ & 20.88
    \\[1.5mm]
    D     &  4\,TeV &  1.6\,TeV & 1.6\,TeV & $3\!\times\!10^{-2}$   & $2\!\times\!10^{-2}$       & 2670
    \\[1.5mm]
   \hline\hline
 \end{tabular}
 \caption{Four samples (A,\,B,\,C,\,D) of our parameter set, which lead to successful Higgs inflation.}
 \label{tab:2}
 \end{table}

  Given these new ingredients, we are ready to analyze the renormalization group running
  for couplings and fields.  We will use the full set of $\beta$-functions above the threshold of
  heavy particles, and the SM $\beta$-functions below the threshold.
  To link these different regions,
  we integrate out $\,\SS\,$ at $\,\mS\,$ and $\,\TT\,$ at $\,\mT\,$,\,
  by inserting their equations of motion into the Lagrangian. This will impose the matching condition
  $\,\lam_1^{}=\lam + \lam_3^2/(4\lam_2^{})\,$ for the scalar threshold $\,\mu=\mS\,$,\,
  and $\,y_1^{}=y_t^{}\,$  for the fermion threshold $\,\mu=\mT\,$,\,
  where $\,\lam\,$ and $\,y_t^{}\,$  are the Higgs self-coupling and top-Yukawa coupling of the SM,
  respectively.
  In addition, we also use the matching condition at $\,\mu=\mt\,$
  as described in \cite{StabilityNNLO}. For the nonminimal coupling $\,\xi\,$,\,
  one may set its initial value at some high scale.
  As shown in Table\,\ref{tab:2}, our Samples (A,\,B,\,C) have $\,\xi =\order{1-20}$\,
  which respects perturbative unitarity \cite{XRH}.
  So it is fine to set the initial value of $\,\xi\,$ at the Planck scale $\,\Mp$\,.\,
  For the Sample-D with $\,\xi=\order{10^3}\,$,\,
  we set the initial value of $\,\xi\,$ at $\,\mu = \Mp/\xi\,$.

 \begin{figure}
   \centering
   \includegraphics[height=6.8cm,width=0.48\textwidth]{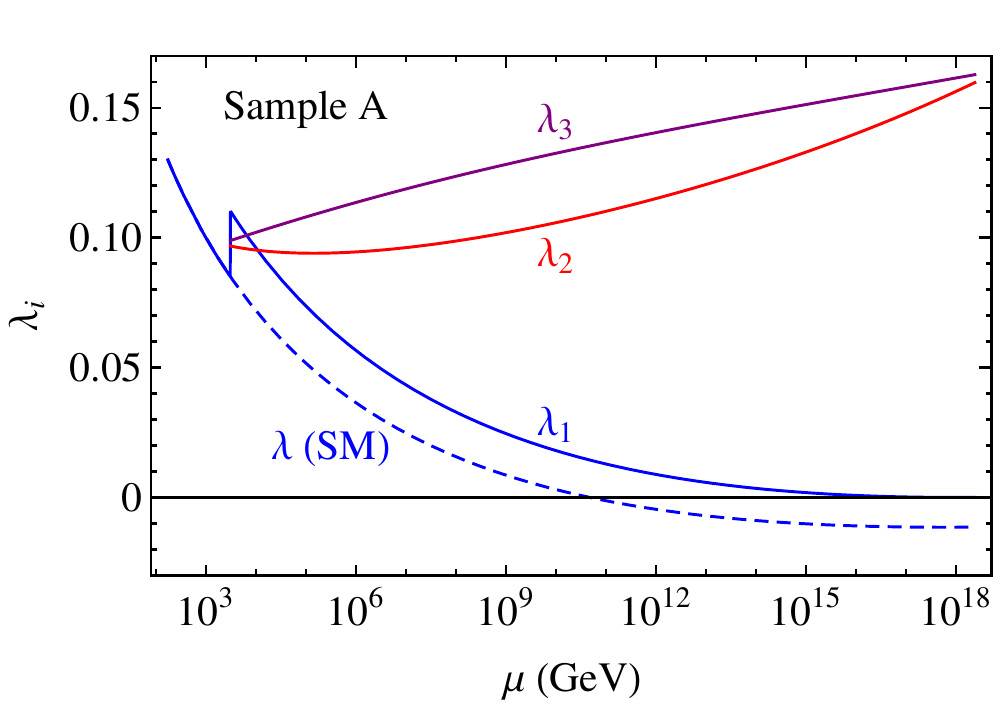}
   \includegraphics[height=6.8cm,width=0.48\textwidth]{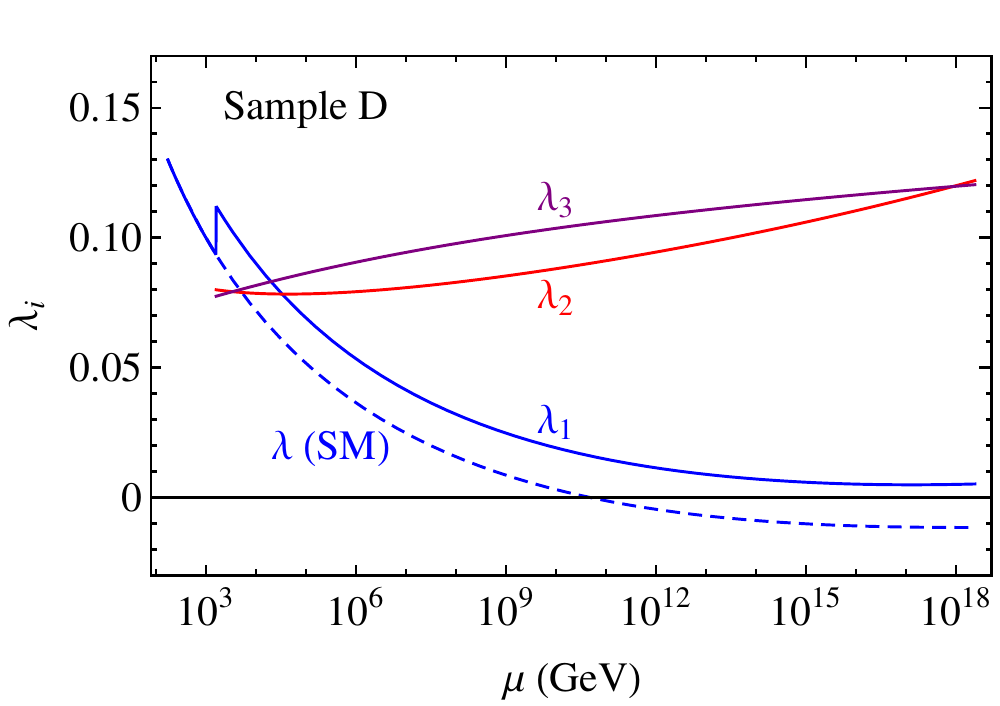}
   \caption{Running scalar couplings $\,(\lam_1^{},\,\lam_2^{},\,\lam_3^{})$\,
   as functions of the energy scale $\,\mu$\,,\, from bottom to top.
   The Sample-A and Sample-D, as defined in Table\,\ref{tab:2},
   are used for the left and right plots, respectively.
   For comparison, we also depict the running of the SM Higgs coupling
   $\,\lam\,(\text{SM})\,$ in each plot by the blue dashed curve.
   }
   \label{fig:1}
 \end{figure}

 For numerical analysis, we input the Higgs mass and top mass to be about their
 current experimental central values,
 $\,m_h^{}=125.6$\,GeV \cite{LHC2013} and $\,\mt =173.3$\,GeV \cite{mt-exp}.
 In Table\,\ref{tab:2}, we have constructed four representative samples
 (A,\,B,\,C,\,D) for the model-parameters.
 As will be demonstrated below, all the four samples lead to successful Higgs inflation. In our analysis,
 we input both top and Higgs masses by their experimental central values\,\cite{LHC2013,mt-exp}
 without fine-tuning.  There are some remaining tunings only for the two theory parameters,
 the mixing angle $\,\theta\,$ and the nonminimal coupling $\,\xi\,$ in the critical point scenario,
 corresponding to the Sample-(A,\,B,\,C); while no tuning is needed for the large-$\xi$ scenario
 in our Sample-D.
 In each sample, for simplicity we choose equal masses for the two singlets $\,(\SS,\,\TT)$\,
 at the TeV scale, while in practice they are allowed to have different masses when needed.
 In Fig.\,\ref{fig:1}, we plot the three running couplings $\,(\lam_1^{},\,\lam_2^{},\,\lam_3^{})$\,
 of the scalar sector for Sample-A and Sample-D.
 For comparison, we further plot the SM Higgs self-coupling $\,\lam\,$ with two-loop running,
 shown by the blue dashed curve in each plot of Fig.\,\ref{fig:1}.
 As mentioned in Sec.\,\ref{sec:1}, the SM Higgs self-coupling becomes negative around $\,10^{11}$\,GeV,
 which is far below the inflation scale $\,\Lambda_{\text{INF}}^{} = \order{10^{16}}$\,GeV.
 But, our Fig.\,\ref{fig:1} demonstrates that,
 after including the two new particles $(\SS,\,\TT)$ at TeV scale,
 the Higgs coupling, now called $\,\lam_1^{}\,$,\, is lifted up
 at the mass-threshold $\,\mu =\mS\,$,\,
 and reaches its minimum of $\,\order{10^{-6}}$\, around the Planck scale $\,\Mp^{}$\,.\,
 Such a small $\,\lam_1^{}\,$ can generate a rather flat scalar potential,
 and thus leads to successful inflation.
 In addition, we also plot the quartic scalar couplings $\,(\lam_{2}^{},\,\lam_{3}^{})\,$
 in Fig.\,\ref{fig:1}, to make sure that all the scalar couplings
 are consistent with the stability and perturbativity.
 In our analysis, we have used the RG equations up to two-loop for the SM sector, and one-loop
 for the new physics sector (cf.\ Appendix\,\ref{App}).

 \begin{figure}
   \centering
   \includegraphics[height=6.5cm,width=0.49\textwidth]{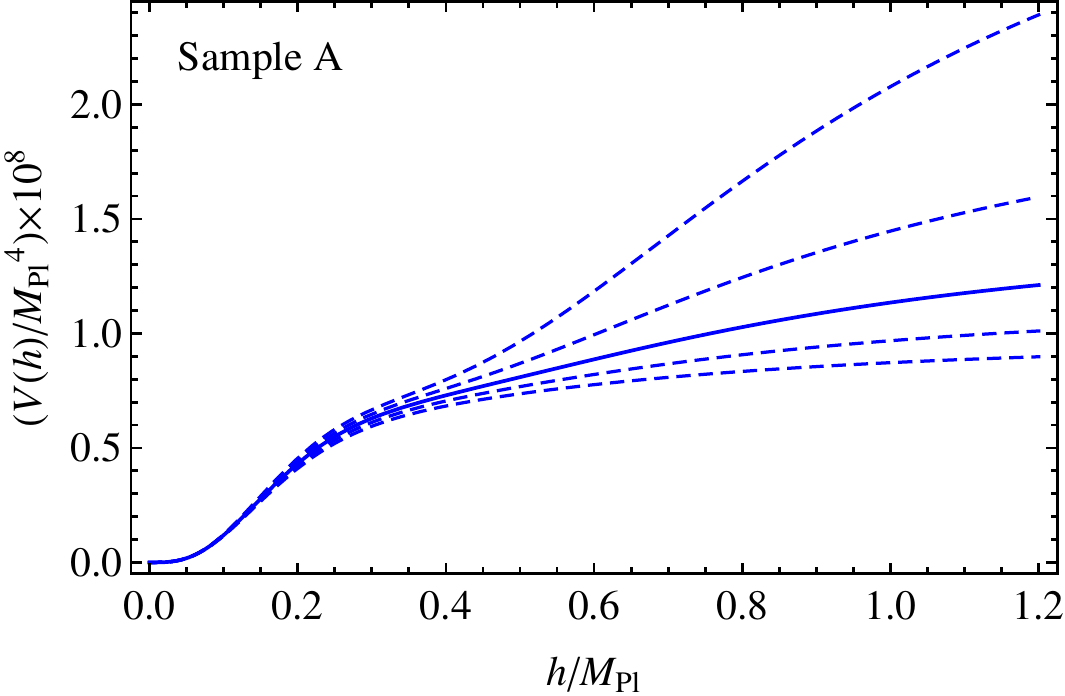}
   \includegraphics[height=6.5cm,width=0.49\textwidth]{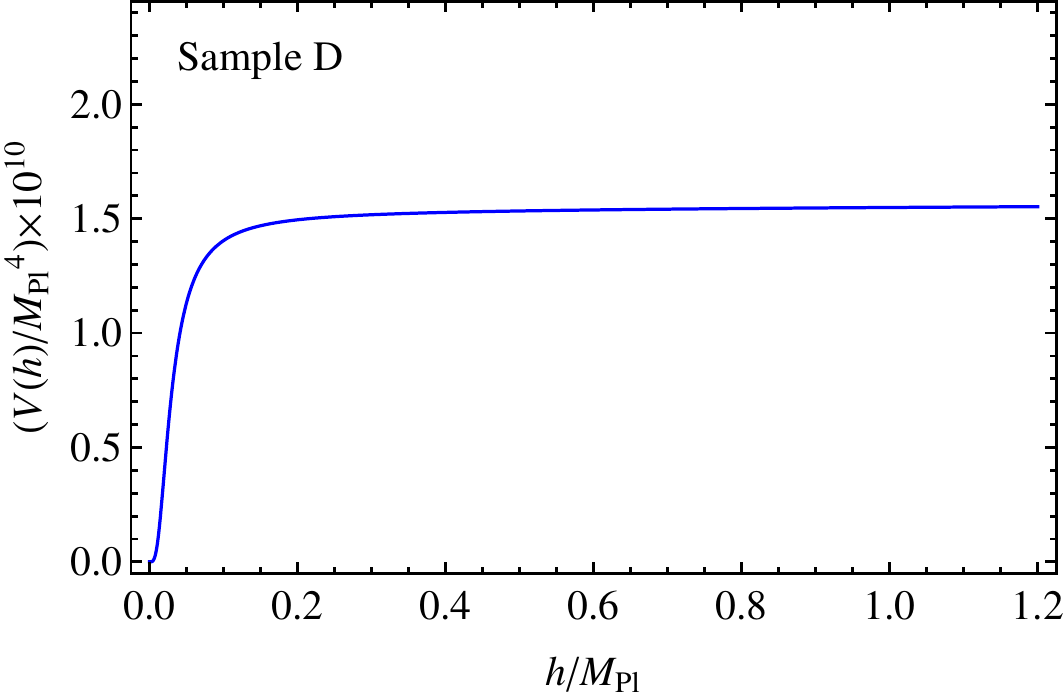}
   \caption{Scalar potential $\,V(h,\SS)$\, in the $\,h\,$ direction.
   The left plot depicts Sample-A.
   From bottom to top, the five curves correspond to the nonminimal coupling
   $\,\xi=\xi_0^{}+\Delta\xi\,$ with $\,\Delta\xi=2,\,1,\,0,\,-1,\,-2$\,,\, respectively,
   where $\,\xi_0^{}\,$ and all other parameters are taken from Sample-A of Table\,\ref{tab:2}.
   The solid curve in the middle describes the potential with a successful Higgs inflation.
   For comparison, the right plot displays Sample-D with a large $\,\xi\,$ from Table\,\ref{tab:2}.}
   \label{fig:2}
 \end{figure}

  Given the running scalar couplings in Fig.\,\ref{fig:1},
  we can compute the scalar potential (\ref{EffPotential}).
  This is shown by the solid curves in Fig.\,\ref{fig:2}, for Sample-A and Sample-D.
  It is clear from these two plots that the scalar potential along $\,h\,$ direction displays
  a nearly flat shape, which also has the proper height to create the observed amplitude
  of curvature perturbation as we will show below.
  To make sure this inflation potential is stable against perturbation along the $\SS$-direction,
  we plot in Fig.\,\ref{fig:3} the scalar potential $\,V(h,\SS)$\, by including nonzero ${\SS}$ field,
  for Sample-A (left panel) and Sample-D (right panel).
  This plot shows that the inflation occurs along the bottom of the potential valley with
  $\,\SS =0\,$,\, which justifies our early setup of $\,\SS =0\,$ in the analysis.
  Here we do not include the wave function renormalization of $\,\SS\,$ field
  since it is only a tiny correction and irrelevant to our calculation of inflation variables.
  Because the other scalar couplings $\,(\lam_{2}^{},\,\lam_3^{})\,$
  remain positive and perturbative during
  the whole process of inflation, it is evident that the inflation path along $h$-direction
  is stable, as clearly shown in Fig.\,\ref{fig:3}.
  The same conclusion can be drawn for Sample-(B,\,C), where the shape of
  the scalar potential is nearly the same as that of Sample-A.

  To make predictions for Higgs inflation, we compute the first two slow-roll parameters
  $\,\ep\,$ and $\,\eta\,$,
  \beqs
  \beqa
   \ep  &\!=\!\!&
   \FR{\Mp^2}{2}\FR{V'^2_\chi}{V^2}
   \,=\, \FR{\Mp^2}{2}\(\!\FR{\di h}{\di\chi}\!\)^{\!\!2}\FR{V'^2_h}{V^2} \,,
  \\[1mm]
   \eta &\!=\!\!&
   {\Mp^2}\FR{V''_\chi}{V} \,=\, \FR{\Mp^2}{V}\FR{\di h}{\di\chi}\FR{\di}{\di h}
   \(\!\FR{\di h}{\di\chi}V'_h\!\).
  \eeqa
  \eeqs
 The inflation ends whenever $\,\ep\simeq 1\,$ or $\,|\eta|\simeq 1\,$
 (corresponding to $\,h=h_{\text{end}}^{}$),\,
 before which the universe experiences a period of nearly exponential expansion.
 The total amount of inflation can be quantified by the number of $e$-foldings $\,N_e\,$,\,
 which can be derived from the scalar potential,
 \beqa
   N_e ~=\, \FR{1}{\,\Mp^2\,}\int_{h_{\text{end}}^{}}^{h_0^{}}\!\!\di h\,
   \(\!\FR{\di\chi}{\di h}\!\)^{\!\!2}\FR{V}{V'_h} \,.
 \eeqa
 The required value of $\,N_e\,$ for observable inflation depends on the process of reheating.
 An analysis of reheating in Higgs inflation gives roughly
 $\,N_e\simeq 59\,$ \cite{Reheating}.
 Then, we can evaluate the slow-roll parameters at the beginning of these 59 folds of inflation,
 namely, at $\,h=h_0^{}\,$,\, to get the predictions for the spectral index
 $\,n_s^{}=1-6\ep+2\eta$\, and the tensor-to-scalar ratio $\,r=16\,\ep\,$.\,
 Furthermore, we need to make sure that the observed amplitude of curvature perturbation
 $\,V/\ep\simeq (0.027\Mp^{})^4\,$ \cite{PlanckCosPara} is appropriately produced.

 \begin{figure}
   \centering
   \includegraphics[width=11.1cm,height=9.1cm]{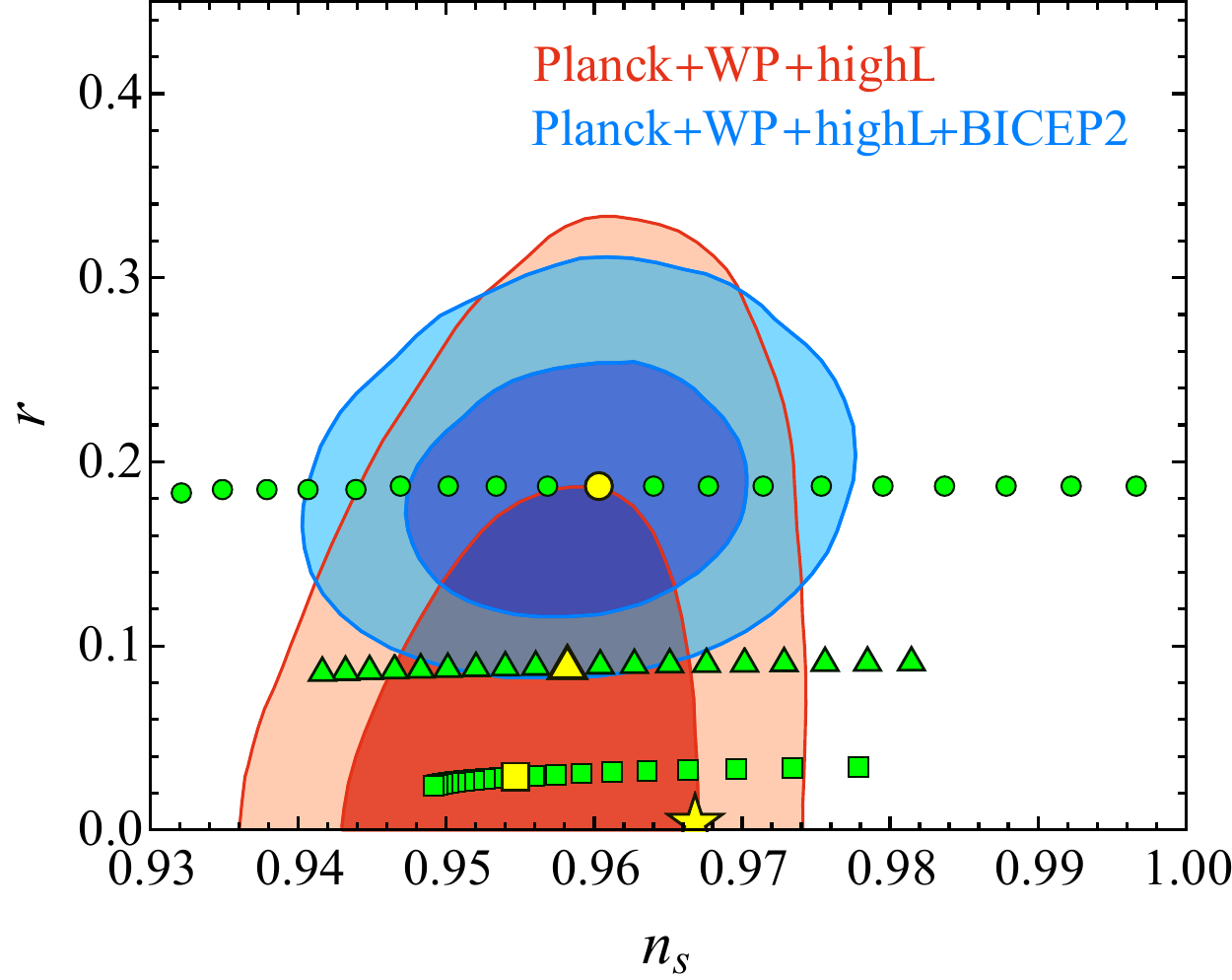}
   \vspace*{-2mm}
   \caption{Sample predictions of our model for the spectral index $\,n_s^{}\,$
   and tensor-to-scalar ratio $\,r\,$.\, The yellow (round, triangular, square, star) dot corresponds to
   Sample-(A,\,B,\,C,\,D) in Table\,\ref{tab:2}. The green (round, triangular, square) dots depict the
   predictions of varying the nonminimal coupling $\,\xi\,$
   within $\,(\pm 0.0005,\,\pm 0.002,\,\pm 0.05)\,$ for Sample-(A,\,B,\,C), respectively.
   The shaded regions are observed limits at 68\%\,C.L. and 95\%\,C.L.,
   taken from Fig.\,13 of Ref.\,\cite{BICEP2},
   where the measurement was made for $k=0.002$\,Mpc$^{-1}$.\,}
   \label{fig:4}
 \end{figure}

 Next, we take the four sets of sample parameters in Table\,\ref{tab:2} to compute these observables.
 For Sample-A, we derive the amplitude of curvature perturbation
 $\,(V/\ep)^{1/4}\simeq 0.027\Mp^{}\,$ at $\,h_0^{}\simeq 0.866\Mp^{}\,$.\,
 For the spectral index and the tensor-to-scalar ratio, we deduce
 \beqa
 \label{eq:A-ns-r}
 (n_s^{},\,r) ~\simeq~ (0.960,\,0.186), && \hspace*{8mm} (\text{Sample-A}).
 \eeqa
 Accordingly, for Sample-B, we compute the amplitude of curvature perturbation
 $\,(V/\ep)^{1/4}\simeq 0.028\Mp^{}$\, at $\,h_0^{}\simeq 0.819\Mp^{}\,$.\,
 We further derive spectral index and the tensor-to-scalar ratio,
 \beqa
 \label{eq:B-ns-r}
 (n_s^{},\,r) ~\simeq~ (0.958,\,0.091), && \hspace*{8mm} (\text{Sample-B}).
 \eeqa
 For Sample-C, we infer
 $\,(V/\ep)^{1/4}\simeq 0.028\Mp^{}$\, at $\,h_0^{}\simeq 0.731\Mp^{}\,$,\, and
 \beqa
 \label{eq:C-ns-r}
 (n_s^{},\,r) ~\simeq~ (0.955,\,0.028), && \hspace*{8mm} (\text{Sample-C}).
 \eeqa
 For Sample-D, we deduce  $\,(V/\ep)^{1/4}\simeq 0.028\Mp^{}$
 at $\,h_0^{}\simeq 0.178\Mp^{}\,$,\, and
 \beqa
 \label{eq:D-ns-r}
 (n_s^{},\,r) ~\simeq~ (0.967,\,0.005), && \hspace*{8mm} (\text{Sample-D}).
 \eeqa
 We present these predictions in Fig.\,\ref{fig:4}, where the results of
 Sample-(A,\,B,\,C,\,D) are denoted by the yellow (round, triangular, square, star) dots,
 respectively.
 In the same figure, we also plot the predicted values of $\,(n_s^{},\,r)$\,
 by varying the nonminimal coupling $\,\xi\,$ up to
 $\,\Delta\xi_{\max}^{} = (\pm 0.0005,\, \pm 0.002,\, \pm 0.05)$\,
 from the $\,\xi\,$ values of Sample-(A,\,B,\,C)
 with a step equal to $\,0.1\Delta\xi_{\max}^{}\,$,\,
 which are marked by green (round, triangular, square) dots for Sample-(A,\,B,\,C).
 It is clear from Fig.\,\ref{fig:4} that our model can successfully produce a period of inflation
 in the early universe. For instance, the predicted observables of Sample-(A,\,B)
 agree with the (combined) BICEP2 data \cite{BICEP2}, while the Sample-(B,\,C,\,D)
 have good fit with the Planck+WMAP+highL data \cite{Planck}.

\vspace*{2mm}
\section{Conclusions and Discussions}
\label{sec:4}
\vspace*{2mm}

 Higgs inflation is among the most economical and predictive inflation models on the market,
 although it has tension with the collider measurements of Higgs and top quark masses.
 In this work, we constructed a minimal extension of the original Higgs inflation
 \cite{HI1,HI2,HIF-rev}\cite{Allison,HIcrit1,HIcrit2}
 by adding only two weak-singlet particles at TeV scale,
 a real scalar $\,\SS\,$ and a vector-quark $\,\TT$\,.\,
 In Sec.\,\ref{sec:2}, we first explained why we
 need to include two new weak-singlets instead of one for constructing a minimal extension of
 the Higgs inflation. Then, we presented our model in Table\,\ref{tab:1} and
 Eqs.\,\eqref{ScalarPotential},\eqref{eq:L-Yukawa},\eqref{eq:NMC}.
 From these, we derived the mass-spectra, mixing angles and couplings for both the scalar sector
 and the quark sector. In Sec.\,\ref{sec:3}, we demonstrated that this minimal extension leads
 to successful Higgs inflation, consistent with the observations of BICEP2 and/or Planck,
 as well as the collider measurements on top and Higgs masses.
 In particular, we explicitly constructed representative Samples (A,\,B,\,C,\,D)
 as in Table\,\ref{tab:2}, and presented their running scalar couplings and
 the shape of scalar potentials in Figs.\,\ref{fig:1}-\ref{fig:3}.
 We further derived the predicted spectral index $\,n_s^{}\,$ and tensor-to-scalar ratio $\,r\,$
 in Eqs.\,\eqref{eq:A-ns-r}-\eqref{eq:D-ns-r}. In Fig.\,\ref{fig:4}, we made explicit comparison of our
 sample predictions of $\,(n_s,\,r)$\, with the measurements from BICEP2 \cite{BICEP2}
 and Planck \cite{Planck}.

 Some discussions are in order. A nice feature is that the present model does not
 finely tune the top and Higgs masses, and thus provides a better realization
 than the previous models of Higgs inflation \cite{HIcrit1,HIcrit2}.
 We note that for realizing $\,r=\order{0.1}\,$ in Sample-(A,\,B),
 there is some remaining tuning on the two theory parameters
 (the $t-\TT$ mixing angle $\theta$ and the nonminimal coupling $\xi$).
 This is expected since the flatness of the scalar potential
 during inflation is achieved almost solely by the renormalization group running of scalar coupling,
 and thus is rather sensitive to the choice of initial conditions.
 But, instead of treating the fine-tuning as a problem, we may consider
 it as a nontrivial constraint on the model from cosmology data,
 and this constraint can be directly tested at the LHC and future collider experiments.
 For instance, the mixing between Higgs boson $\,h\,$ and the heavy scalar $\SS$, as well as
 the mixing between top quark $t$ and heavy vector-quark $\TT$,
 will modify both the production rate and the decay width of the Higgs boson
 $\,h\,(125\,\text{GeV})\,$
 at the LHC. Our analysis shows that such mixings are fairly small, around the $\order{10^{-2}}$,\,
 and are thus fully consistent with the current LHC data so far.
 Since the vector-quark $\TT$ joins
 QCD interactions and has mass around $1-3$\,TeV, we expect it can be directly produced at the
 upcoming LHC\,(14\,TeV) runs by gluon fusions $\,gg\to\TT\,\over{\TT}\,$,\,
 and it mainly decays via $\,\TT\to Wb\,$ due to the $\,t-\TT\,$ mixing.
 The LHC\,(14\,TeV) runs could produce the heavy scalar $\,\SS\,$ via gluon fusion channel
 $\,gg\to \SS\,$ with the subsequent decays\footnote{The production and decays of an extra
 heavier neutral scalar (which mixes with the observed 125\,GeV Higgs particle $\,h\,$)
 was studied before for the LHC in different model contexts \cite{Wang:2013jwa}.}
 $\,\SS\to WW,ZZ,hh\,$, which may be detected via
 $\,WW\to 2\ell 2\nu\,$,\, $\,ZZ\to 4\ell\,$,\,
 and $\,hh\to (WW^*)(b\bar{b})\to (2\ell 2\nu)(b\bar{b})\,$,\,
 or $\,hh\to (\gamma\gamma)(b\bar{b})\,$,\, etc.
 The future high energy circular $pp$ colliders ($50-100$\,TeV)\,\cite{FCC}
 should have much better chance
 to discover such TeV-scale heavy singlets $\,\TT\,$ and $\,\SS\,$.\,

 In our analysis, we have chosen the renormalization scale in the Einstein frame according
 to prescription-I of \cite{Bezrukov:2009db}.
 One may also consider the alternative scenario in Jordan frame,
 which would correspond to the chaotic inflation with quadratic potential \cite{HIcrit1}.
 In addition, it is useful to explore more systematically the full viable parameter space of this model,
 which may have reduced tuning.  This will also give wider ranges of the
 couplings and masses of $\,(\TT,\,\SS)$,\, which are useful for collider searches of such TeV scale
 vector-quark and neutral scalar.
 Finally, we may also consider embedding of our minimal extension into a SUSY framework.

\vspace*{5mm}
\addcontentsline{toc}{section}{Acknowledgments\,}
\noindent
{\bf\large Acknowledgements}\\[1.5mm]
 We thank John R.\ Ellis, Yuta Hamada, Josh Ruderman, and Alexander Spencer-Smith
 for useful discussions on this subject.  We also thank Michelangelo Mangano and
 Michael E.\ Peskin for discussing the top mass measurements.
 This work was supported by National NSF of China (under grants 11275101, 11135003)
 and National Basic Research Program (under grant 2010CB833000).


\appendix

\noindent

\vspace*{7mm}
\section{Renormalization Group Equations for Higgs Inflation Analysis}
\label{App}
\vspace*{2mm}

{\allowdisplaybreaks

 In this appendix, we summarize the renormalization group equations
 which we have used to solve the running couplings.
 For the present model, we have three SM gauge couplings $(g_s^{},\,g,\,g')$
 for the SM gauge group $SU(3)_c\otimes SU(2)_L\otimes U(1)_Y$,\,
 three scalar couplings $(\lam_1,\,\lam_2,\,\lam_3)$ for the Higgs potential,
 three Yukawa coupling $(y_1^{},\,y_2^{},\,y_3^{})$,\, and a nonminimal coupling $\xi$.\,
 They obey the following renormalization group equations,
 \bge
   \FR{\di X}{\,\di \ln\mu\,} ~=~ \beta(g_i^{},\,\lam_i^{},\,y_i^{},\,\xi) \,,
 \ede
 where $\,X\,$ represents any coupling listed above.
 In the following, we will present all the relevant $\beta$ functions needed
 for computing the scalar potential.

 We first summarize the $\beta$ functions in the SM
 up to two-loop order with appropriate $s$-insertion \cite{Allison,sInsertion},
 as well as the one-loop $\beta$ function of the nonminimal-coupling $\,\xi$\,,
 \beqs
 \begin{align}
   \be_{g_s}^{}
   =&~\FR{g_s^3}{(4\pi)^2}\Big(-7\Big)+\FR{g_s^3}{(4\pi)^4}\Big(\FR{11}{6}g'^2+\FR{9}{2}g^2-26g_s^2-2sy_1^2\Big),
   \\[2mm]
   \be_{g}^{}
   =&~\FR{g^3}{(4\pi)^2}\Big(-\FR{39-s}{12}\Big)+\FR{g^3}{(4\pi)^4}
   \Big(\FR{3}{2}g'^2+\FR{35}{6}g^2+12g_s^2-\FR{3}{2}sy_1^2\Big),
   \\[2mm]
   \be_{g'}^{}
   =&~\FR{g'^3}{(4\pi)^2}\Big(\FR{81+s}{12}\Big)+\FR{g'^3}{(4\pi)^4}
   \Big(\FR{199}{18}g'^2+\FR{9}{2}g^2+\FR{44}{3}g_s^2-\FR{17}{6}sy_1^2\Big),
   \\[2mm]
   \be_{\lam_1^{}}^{}
   =&~\FR{1}{(4\pi)^2}\Big(6(1+3s^2)\lam_1^2-6y_1^4+\FR{3}{8}
   \big(2g^4+(g^2+g'^2)^2\big)+\lam_1(-9g^2-3g'^2+12y_1^2)\Big)\n\\
                &~+\FR{1}{(4\pi)^4}\bigg[\FR{1}{48}\Big((912+3s)g^6-(290-s)g^4g'^2-(560-s)g^2g'^4-(380-s)g'^6\Big)
   \n\\[1.5mm]
                &~+(38-8s)y_1^6-y_1^4\Big(\FR{8}{3}g'^2+32g_s^2+(12-117s+108s^2)\lam_1\Big)\n\\
                &~+\lam_1\Big(-\FR{1}{8}(181+54s-162s^2)g^4+\FR{1}{4}(3-18s+54s^2)g^2g'^2
                +\FR{1}{24}(90+377s+162s^2)g'^4
                \n\\[1.5mm]
                &~+(27+54s+27s^2)g^2\lam_1+(9+18s+9s^2)g'^2\lam_1
                -(48+288s-324s^2+624s^3-324s^4)\lam_1^2\Big)
                \n\\[1.5mm]
                &~+y_1^2\Big(-\FR{9}{4}g^4+\FR{21}{2}g^2g'^2-\FR{19}{4}g'^4+\lam_1
                \big(\FR{45}{2}g^2+\FR{85}{6}g'^2+80g_s^2-(36+108s^2)\lam_1\big)\Big)\bigg],
   \\[2mm]
   \be_{y_1^{}}^{}
   =&~\FR{y_1^{}}{(4\pi)^2}
      \left[-\FR{9}{4}g^2-\FR{17}{12}g'^2-8g_s^2+\(\FR{23}{6}+\FR{2}{3}s\)y_1^2\right]
   \n\\[1.5mm]
             &~+\FR{y_1}{(4\pi)^4}\bigg[-\FR{23}{4}g^4-\FR{3}{4}g^2g'^2
             +\FR{1187}{216}g'^4+9g^2g_s^2+\FR{19}{9}g'^2g_s^2-108g_s^4
             \n\\[1.5mm]
             &~+\(\FR{225}{16}g^2+\FR{131}{16}g'^2+36g_s^2\)s y_1^2
               +6\(-2s^2y_1^4-2s^3y_1^2\lam_1+s^2\lam_1^2\)\bigg],
   \\[2mm]
   \be_\xi^{}
   =&~\FR{\,\xi+\!1/6\,}{(4\pi)^2}
      \left[ -\FR{9}{2}g^2-\FR{3}{2}g'^2+6y_1^2+(6+6s)\lam_1^{}\right] .
 \end{align}
 \eeqs
 Derivation of the beta function $\,\beta_\xi^{}\,$ for the nonminimal coupling $\,\xi\,$
 is reviewed in \cite{RGxi}.

 The additional $\beta$-functions from the beyond SM sector are given as follows.
 We do not include $s$-insertion for these terms as their effects are indirect and negligibly small.
 We also do not include new particle sector on the running of nonminimal coupling $\xi$ for the same reason.
 \beqs
 \begin{align}
   \Delta \be_{g_s^{}}^{} = &~\FR{g_s^3}{(4\pi)^2}\FR{2}{3}\,,
   \hspace{5mm}
   \Delta \be_{g}^{}  \,=\, 0 \,,\hspace{5mm}
   \Delta \be_{g'}^{} \,=\, \FR{g'^3}{(4\pi)^2}\FR{16}{9} \,,
   \\[2mm]
   \Delta\be_{\lam_1^{}}^{} =&~\FR{1}{(4\pi)^2}\Big(\FR{1}{2}\lam_3^2+12\lam_1y_2^2-6y_2^4-12y_1^2y_2^2\Big),
   \\[2mm]
   \Delta\be_{y_1^{}} = &~\FR{9}{2(4\pi)^2}y_1^{} y_2^2 \,,
   \\[2mm]
   \be_{\lam_2^{}}^{} = &~\FR{1}{(4\pi)^2}\Big(18\lam_2^2+12y_3^2\lam_2+2\lam_3^2-6y_3^4\Big),
   \\[2mm]
   \be_{\lam_3^{}}^{} =
   &~\FR{1}{(4\pi)^2}\!\left[\lam_3^{}\big( 12\lam_1^{}\!+6\lam_2^{}\!+4\lam_3^{}\!
      +6y_1^2\!+6y_2^2\!+6y_3^2\!
     -\FR{9}{2}g^2\! -\FR{3}{2}g'^2\big)\! - 12y_2^2y_3^2\right] \!,
   \\[2mm]
   \be_{y_2^{}} =&~ \FR{y_2}{(4\pi)^2}\(\FR{9}{2}y_1^2+\FR{9}{2}y_2^2-8g_s^2-\FR{9}{4}g^2-\FR{17}{12}g'^2\)\!,
   \\[2mm]
   \be_{y_3^{}}=&~\FR{y_3}{(4\pi)^2}\(y_2^2+\FR{9}{2}y_3^2-8g_s^2-\FR{8}{3}g'^2\) \!.
 \end{align}
 \eeqs

}


\vspace{10mm}
%

\end{document}